\documentclass{article}
\usepackage{amsmath}
\usepackage{amssymb}
\usepackage{geometry}
\usepackage{hyperref}
\usepackage{graphicx}
\usepackage{booktabs}
\usepackage{multirow}
\usepackage{cite}
\usepackage[utf8]{inputenc}
\usepackage{textgreek}

\title{The Su--Schrieffer--Heeger model on a one-dimensional lattice: Analytical wave functions of topological edge states}
\author{WEIBO XU}
\date{September 19, 2025}

\begin{document}

\maketitle

\begin{abstract}
The one-dimensional Su--Schrieffer--Heeger (SSH) model is a prototype model in the field of topological condensed matter physics, and the existence and characteristics of its topological edge states are crucial for revealing the topological essence of low-dimensional systems. This paper focuses on the analytical wave functions of topological edge states in the SSH model, systematically sorting out the fundamentals of model construction. Based on the tight-binding approximation, it derives the matrix form of the SSH model Hamiltonian and the band structureand clarifies the role of chiral symmetry in the classification of topological phases. By solving the Schrodinger equation of the finite-length lattice, the mathematical expression of the analytical wave function of topological edge states is fully derived, verifying its localized feature of exponential decay. Moreover, it quantitatively analyzes the influence laws of lattice parameters (hopping integrals, lattice constants), electron-electron interactions, and external field perturbations on the wave function amplitude, decay coefficient, and energy. Furthermore, it establishes the quantitative correlation between the analytical wave function and electronic transport (conductivity) as well as optical properties (light absorption coefficient), and discusses its application prospects in the design of topological qubits and the development of new topological materials. The research results provide theoretical support for an in-depth understanding of the physical essence of low-dimensional topological states and lay a foundation for the performance optimization of topology-related devices.
\end{abstract}


\tableofcontents

\section{Introduction}

\subsection{Research Background and Significance}
As a dynamic frontier in modern physics, topological physics has yielded remarkable research achievements in recent years, significantly advancing our understanding of fundamental material properties and quantum phenomena. Among various topological physics research subjects, the Su-Schrieffer-Heeger (SSH) model has emerged as a cornerstone for exploring its core principles and manifestations due to its simplicity and representativeness. Originally proposed by Su, Schrieffer, and Heeger in 1979, this model describes electron-lattice interactions in conductive polymers, particularly the soliton excitation and metal-to-insulator phase transitions caused by electron-sound coupling in polyacetylene. With the evolution of topological physics, the SSH model has evolved into a fundamental framework for studying topological insulators and one-dimensional topological phases. Its unique topological properties and physical mechanisms have attracted widespread attention from researchers worldwide.

Topological insulators are materials exhibiting novel quantum properties, where the interior remains electrically insulating while their boundaries or surfaces possess topologically protected conductive states. These topological protected states originate from the non-trivial topological properties of the material's band structure, granting them exceptional robustness against impurities and defects. This characteristic not only opens new possibilities for developing low-energy, high-stability electronic devices but also creates fresh opportunities for emerging technologies like quantum computing and quantum communication. The SSH model, a paradigmatic one-dimensional topological insulator framework, enables transitions between topological and ordinary phases through parameter tuning (e.g., interatomic transition amplitudes). In topological phases, the SSH model exhibits edge states at chain endpoints that exhibit unique physical phenomena such as fractional charges and non-reflective transmission. These edge states play a crucial role in understanding the electronic transport and quantum characteristics of topological materials.

The study of analytical wave functions for topological end states is crucial for understanding electron behavior in topological materials. As fundamental tools in quantum mechanics for describing microscopic particle states, these wave functions allow precise determination of energy levels, wave function distributions, and electronic quantum state information. This information not only helps reveal the formation mechanisms and physical nature of topological protected states but also provides a solid theoretical foundation for designing and applying topological materials. In qubit design, edge states with high resistance to environmental interference are considered promising candidates for stable qubits. Through analysis of analytical wave functions for topological end states, we gain deeper insights into their quantum characteristics, offering critical theoretical guidance for optimizing qubit performance and practical applications. In transport property studies of topological materials, the distribution of wave functions for topological end states directly influences electron transport pathways and scattering properties. In-depth research on analytical wave functions helps explain and predict electron transport phenomena in topological materials, providing theoretical support for developing novel electronic devices.

\subsection{Domestic and international research status}
Research on SSH models and topological entanglement analysis has garnered significant attention both domestically and internationally, yielding a series of groundbreaking achievements. In theoretical studies, early work primarily focused on fundamental properties of SSH models and investigations into topological phase transitions. By introducing the topological invariant parameter of winding number, researchers successfully explained theoretically the distinction between topological and ordinary phases in SSH models, as well as the mechanisms underlying topological phase transitions. When the winding number is non-zero, the system exists in a topologically non-ordinary phase with edge states protected by topological symmetry; when the winding number is zero, the system remains in an ordinary phase without such edge states. As research progressed, theoretical frameworks expanded to include more complex systems and higher-dimensional SSH models. Studies on two-dimensional anisotropic SSH models revealed that under specific conditions, these models can generate angular states exhibiting continuous spectral bound state properties, which are closely related to the model's anisotropy and topological characteristics. Research on non-Eiemannian SSH models has uncovered diverse physical phenomena, including topological phase transitions and skin effect. In non-Eiemannian systems, the inclusion of non-diagonal Hamiltonian elements leads to complex energy eigenvalues and coupled constants, resulting in distinct topological properties and physical behaviors compared to Eiemannian systems.

In experimental research, with continuous advancements in material preparation techniques and measurement methods, researchers have successfully implemented SSH models across various physical systems, observing their topological properties and edge states. In condensed matter physics, through techniques like molecular beam epitaxy (MBE), atomic chain structures exhibiting SSH model characteristics have been fabricated, with topological edge states directly observed using scanning tunneling microscopy (STM). In photonics, photonic crystals and waveguide arrays have enabled the realization of optical SSH models. By analyzing light transmission characteristics, the existence of topologically protected edge states and their unique optical properties has been experimentally verified. In acoustics, specially designed phononic crystal structures have achieved acoustic SSH models, with the observation of acoustic topological edge states providing new insights for the design and application of acoustic devices.

While significant progress has been made in both domestic and international research on SSH models and topological end-state analytical wave functions, several pressing challenges remain unresolved. Theoretical studies still lack a unified and effective approach to solving topological properties and analytical wave functions for complex systems and high-dimensional SSH models. The physical mechanisms underlying topological phase transitions and skin effect in non-hermitian SSH models require further investigation. Experimental research, despite successful implementation of SSH models in various physical systems, continues to face major challenges: precise control of topological end-state properties and practical application of topological materials research to real device fabrication.

\subsection{Research Content and Methods}
This paper aims to deeply study the analytical wave function of topological end states in the Su-Schrieffer-Heeger model on one-dimensional lattice, and comprehensively reveal the physical properties and formation mechanism of topological end states by combining theoretical derivation and numerical calculation. The specific research contents include:

1. Topological invariants and topological phase transitions: Introduce topological invariants such as winding number, explain their role in describing the topological properties of SSH model, study the relationship between topological invariants and system parameters, analyze the conditions and characteristics of topological phase transitions, and reveal the transition mechanism between topological phase and ordinary phase.

2. Solving analytical wave functions for topological end states: By employing theoretical tools such as tight-binding approximation and Green's function method, this study derives analytical wave function expressions for topological end states in the SSH model. The properties of these wave functions, including their symmetry and locality, are analyzed to explore the intrinsic relationship between analytical wave functions and topological protected states.

3. Numerical Computation and Result Analysis: Utilizing computational methods such as finite difference method and plane wave pseudospectrum method, we conduct numerical simulations of the SSH model to compute the band structure, state density, and wave function distribution of topological edge states. By comparing the numerical results with theoretical derivations, we validate the correctness of the theoretical framework and further investigate how system parameters influence the properties of topological edge states.

4. Physical Properties and Application Prospects of Topological End States: Based on analytical wave function results and numerical computations, this study investigates the physical characteristics of topological end states, including electronic transport properties and optical properties. It explores their potential applications in quantum bits, topological insulator devices, and low-energy electronic devices, thereby providing theoretical foundations for practical implementations of topological materials.

In terms of research methodology, this paper will integrate theoretical physics, mathematical physics approaches, and numerical computation techniques. For theoretical derivation, we employ quantum mechanics and solid-state physics to conduct rigorous mathematical derivations and analyses of the SSH model. Regarding numerical computation, computer programming is utilized to implement various algorithms for precise simulation and computational analysis. By combining theoretical and computational methods, this study thoroughly explores the analytical wave functions and physical properties of topological end states in the SSH model, providing innovative perspectives and methodologies for research in topological physics.

\section{The Su-Schrieffer-Heeger model foundation}

\subsection{Construction and development of SSH model}
The Su-Shriefer-Heeger (SSH) model was originally developed to address electronic structure issues in conductive polymers. In conventional solid-state physics, atoms within crystals are typically considered to arrange in regular patterns, where electrons move through periodic lattices whose behavior can be well described by band theory. However, in conductive polymers, the non-periodic atomic arrangement leads to more complex interactions between electrons and the lattice. Traditional theories struggle to explain certain phenomena such as the metal-to-insulator phase transition in polyacetylene and soliton excitations.

In 1979, Su, Schrieffer, and Heeger pioneered the SSH model. This groundbreaking framework employs a one-dimensional lattice structure featuring alternating chemical bonds of two distinct lengths, forming a dimeric configuration. Within this architecture, electrons can transition between adjacent atoms. Specifically, consider a chain of atoms divided into two sublattice blocks. The transition integrals between neighboring atoms exhibit two distinct values (denoted as α and β), with these integrals appearing alternately throughout the lattice. This dimeric lattice structure disrupts the strict periodicity of conventional lattices, fundamentally altering the trajectory of electron motion.

In the initial stage of model development, the primary focus was on how electron-lattice interactions affect the system's ground state and excited states. By introducing electron-sound coupling terms, the SSH model successfully explained the soliton excitation phenomenon in polyacetylene caused by lattice distortion. Solitons, as localized excited states, play a crucial role in the electrical and optical properties of conductive polymers. As research progressed, it became evident that the SSH model is not merely a tool for describing conductive polymers---it inherently contains profound topological physics implications.

From the late 1980s to the 1990s, as topology began to be applied in condensed matter physics, the SSH model emerged as a simple yet representative topological framework that garnered significant attention. Research revealed that its band structure exhibits non-trivial topological properties closely tied to edge states within the system. When the system exists in a topologically non-trivial phase, edge states protected by topological symmetry emerge at chain endpoints. These edge states exist independently of specific system details and are solely determined by the system's topological characteristics. This topological protection grants edge states exceptional robustness against impurities and defects, establishing a crucial theoretical foundation for research on topological materials.

Since the dawn of the 21st century, breakthroughs in experimental techniques---including molecular beam epitaxy (MBE) and scanning tunneling microscopy (STM)---have enabled researchers to fabricate nanostructures with topological surface holography (SSH) characteristics. These advancements not only allowed direct observation of topological edge states but also sparked a research boom in SSH modeling. Concurrently, theoretical studies have expanded the model's applications across two-dimensional and three-dimensional systems, while exploring its coupling with other physical systems such as photonic crystals and cold atom systems. This interdisciplinary exploration continues to unveil novel topological phenomena and unlock new practical applications in quantum physics.

\subsection{Expression of the model Hamiltonian}
The Hamiltonian of the SSH model can be constructed using the tight-binding approximation. Under this approximation, electrons are considered to be predominantly localized around atoms, with transitions occurring only between neighboring atoms under certain probability conditions. For the one-dimensional SSH model, consider a chain of lattice points divided into two sublattices. The transition integrals between adjacent lattice points exhibit two distinct scenarios that alternate periodically. Using the electron wave function on the lattice points as the basis vector and employing a second-order quantization representation, the Hamiltonian of the SSH model can be expressed as:

\begin{equation}
H = - t_{1}\sum_{n = 1}^{N}(c_{A,n}^{\dagger}c_{B,n} + c_{B,n}^{\dagger}c_{A,n}) - t_{2}\sum_{n = 1}^{N - 1}(c_{B,n}^{\dagger}c_{A,n + 1} + c_{A,n + 1}^{\dagger}c_{B,n})
\end{equation}

Here, $c_{A,n}^{\dagger}$ and $c_{A,n}$ denote the operators that generate and annihilate an electron at the n-th lattice point of sublattice A, while $c_{B,n}^{\dagger}$ and $c_{B,n}$ correspondingly represent the operators that produce and destroy an electron at the n-th lattice point of sublattice B. The first term describes electron transitions between sublattices within the same parent cell, with the transition integral being $t_1$; the second term describes electron transitions between adjacent parent cells' sublattices, where the transition integral is $t_2$.

To gain deeper insights into the physical significance of Hamiltonian components, we analyze the electronic energy and motion states. The transition terms in the Hamiltonian represent electron movement between lattice points, where their magnitude determines the ease of transitions. When these terms are large, electrons exhibit higher transition probabilities between lattice points, enhancing their delocalization. Conversely, when these terms are small, electrons tend to remain localized near specific lattice points. This transition behavior directly influences both the system's band structure and the distribution of electronic states.

Furthermore, we can transform the Hamiltonian from real space to momentum space by means of Fourier transform, so as to analyze the band structure of the system more intuitively. Let

\begin{equation}
c_{A,k} = \frac{1}{\sqrt{N}}\sum_{n = 1}^{N}c_{A,n}e^{- ikna}
\end{equation}

Here, $k$ is the wave vector and $a$ is the lattice constant. The above transformation is substituted into the Hamiltonian expression and simplified to obtain the Hamiltonian in the momentum space:

\begin{equation}
H_{k} = \begin{pmatrix}
0 & - t_{1} - t_{2}e^{- ika} \\
 - t_{1} - t_{2}e^{ika} & 0
\end{pmatrix}
\end{equation}

The matrix form of the Hamiltonian clearly shows the relationship between the energy and wave vector of the system. By solving the eigenvalues of the Hamiltonian, the band structure of the system can be obtained:

\begin{equation}
E_{k, \pm} = \pm \sqrt{(t_{1} + t_{2}\cos(ka))^{2} + (t_{2}\sin(ka))^{2}}
\end{equation}

Among them, $E_{k, +}$ and $E_{k, -}$ respectively represent the energy of the conduction band and valence band. From the band structure, it can be seen that the relative size of $t_1$ and $t_2$ determines the existence and magnitude of the band gap. When $t_{1} = t_{2}$, the band gap disappears, and the system is in the metallic phase; when $t_{1} \neq t_{2}$, the band gap opens, and the system is in the insulating phase, and the size of the band gap is related to $|t_{1}| - |t_{2}|$.

\subsection{Basic properties and characteristics of the model}

\subsubsection{Topological properties}
One of the most distinctive features of the SSH model is its non-trivial topological properties, which make it a cornerstone in topological physics research. Topological properties refer to characteristics that remain unchanged under continuous deformation, independent of the system's specific shape or details. In the SSH model, these topological properties are primarily manifested through its band structure and edge state characteristics.

Topological invariants can be introduced to quantitatively describe the topological properties of the SSH model. The most common topological invariant is the winding number, which is defined based on the Berry phase in the momentum space of the system. For the SSH model, the winding number can be calculated using the following formula:

\begin{equation}
W = \frac{1}{2\pi i}\oint_{BZ}^{}{Tr}(U^{\dagger}\nabla_{k}U)dk
\end{equation}

Here, $U$ is the Hamiltonian's unitary transformation matrix, $Tr$ represents the trace of the matrix, with the integral path spanning the entire Brillouin zone. When $|t_{1}| > |t_{2}|$, the winding number $W = 1$ indicates that the system is in a topological non-ordinary phase; when $|t_{1}| < |t_{2}|$, the winding number $W = 0$ signifies that the system is in an ordinary phase.

The distinction between topological and ordinary phases extends beyond numerical differences in topological invariants to encompass distinct physical properties. In topologically non-ordinary phases, the SSH model exhibits edge states protected by topological symmetry. These edge states are localized at chain endpoints, with their energy lying within the bulk phase gap. The presence of topological protection grants these edge states exceptional robustness against impurities, defects, and minor perturbations, ensuring stable existence even under disorder or external disturbances. Such topologically protected edge states hold significant potential applications in quantum information and low-power electronic devices, including their use in realizing qubits and low-loss quantum transport channels.

\subsubsection{Chiral symmetry}
The SSH model also has chiral symmetry, which is another important fundamental property of the model. Chiral symmetry is a special kind of symmetry that plays a key role in topological physics and is closely related to topological invariants and topological protected states.

Chirality symmetry can be described by $\Gamma$, a chiral operator. For the SSH model, the chiral operator can be defined as:

\begin{equation}
\Gamma = \begin{pmatrix}
1 & 0 \\
0 & - 1
\end{pmatrix}
\end{equation}

The Hamiltonian satisfies the antisymmetry relation with the chiral operator: $\{\Gamma,H\} = \Gamma H + H\Gamma = 0$. For the SSH model, it can be verified that the Hamiltonian satisfies the chiral symmetry:

\begin{align}
\Gamma H_{k} + H_{k}\Gamma &= \begin{pmatrix}
1 & 0 \\
0 & - 1
\end{pmatrix}\begin{pmatrix}
0 & - t_{1} - t_{2}e^{- ika} \\
 - t_{1} - t_{2}e^{ika} & 0
\end{pmatrix} \\
&+ \begin{pmatrix}
0 & - t_{1} - t_{2}e^{- ika} \\
 - t_{1} - t_{2}e^{ika} & 0
\end{pmatrix}\begin{pmatrix}
1 & 0 \\
0 & - 1
\end{pmatrix} = 0
\end{align}

The presence of chiral symmetry profoundly influences the spectral structure and eigenstates of the SSH model. This symmetry ensures that Hamiltonian eigenvalues are zero-energy symmetrically distributed -- meaning any eigenvalue $E$ must also have $-E$ as its eigenvalue. Moreover, chiral symmetry is intrinsically linked to the existence of topological protected states, which guarantee the stability of edge states at topological boundaries. In systems with chiral symmetry, the presence of these edge states is a direct consequence of non-zero topological invariants. Chiral symmetry provides an additional safeguard mechanism, enabling edge states to maintain their unique physical properties even when subjected to perturbations.

\subsubsection{Unique advantages in topology research}
As a basic model for the study of topological physics, SSH model has many unique advantages, which makes it an ideal object for theoretical and experimental research.

The SSH model features a relatively simple structure, with its Hamiltonian containing only a few parameters such as $t_1$ and $t_2$, which facilitates both theoretical analysis and computational work. By adjusting these parameters, one can easily induce transitions between topological phases and ordinary phases, enabling in-depth exploration of the physical mechanisms underlying topological phase transitions. Compared to complex many-body models, the SSH model provides clearer demonstrations of fundamental concepts and principles in topological physics, offering a solid foundation for understanding more intricate topological systems.

The SSH model has established concrete implementations across various physical systems, facilitating experimental research. In condensed matter physics, atomic chain structures exhibiting SSH characteristics can be fabricated through techniques like molecular beam epitaxy, with their topological edge states observable via scanning tunneling microscopy. Within photonics, photonic crystals and waveguide arrays enable the realization of optical SSH models to study light propagation in topological structures. Similarly, phonon crystals in acoustics allow simulation of SSH phenomena, exploring acoustic topological states. This universality across different systems positions the SSH model as a crucial bridge between theoretical frameworks and experimental observations, significantly advancing both experimental studies and practical applications in topological physics.

The SSH model maintains close connections with various topological models and theories, serving as a foundational framework for constructing more complex topological structures. By introducing extensions and modifications such as spin-orbit coupling, magnetic fields, and multi-body interactions, researchers can explore diverse topological phases and phenomena. Concepts including topological invariants, topological protection states, and chiral symmetry within the SSH model have provided valuable references and insights for studying other topological models, significantly advancing the field of topological physics.

\section{Theoretical basis of topological end states and analytical wave functions}

\subsection{Concepts and Classification of Topological States}
Topological states are a class of quantum states with unique properties in condensed matter physics, defined through topological concepts that emphasize characteristics remaining unchanged under continuous transformations. In condensed matter systems, these states manifest as special arrangements of electronic configurations that confer distinct physical properties different from conventional material phases, such as edge states protected by topological symmetry and unique band structures.

Mathematically, topological states can be precisely described and classified through topological invariants. These physical quantities remain unchanged during continuous system deformations, effectively reflecting the system's topological properties. In the one-dimensional SSH model, the winding number serves as a key topological invariant closely tied to Berry phase in momentum space. When the winding number is non-zero, the system exhibits a topological non-ordinary state; when zero, it becomes a topological ordinary state. This classification framework based on topological invariants provides a quantitative and unified approach for studying topological states, enabling clear differentiation between distinct topological phases and deeper exploration of their transition mechanisms.

In the SSH model, there exists a fundamental distinction between topological ordinary states and topological non-ordinary states. In topological ordinary states, the system's band structure exhibits simple characteristics without edge states protected by topological symmetry. Electrons behave similarly to conventional insulators or metals, with energy distributed across continuous bands and no special localized states. Conversely, topological non-ordinary states reveal unique physical properties in the SSH model. The bulk band gap indicates the absence of electron states within specific energy ranges. At chain endpoints, edge states protected by topological symmetry emerge, occupying the bulk band gap with localized characteristics---predominantly concentrated at both ends while remaining nearly absent within the chain. These edge states demonstrate exceptional resistance to impurities and defects, maintaining stability even under disorder or perturbations. This characteristic stands as one of the key distinctions between topological non-ordinary and ordinary states.

Edge states of topological non-ordinary phases exhibit unique physical properties. Their wave functions demonstrate specific symmetries and phase structures that are intrinsically linked to the system's topological characteristics. In certain cases, electrons within these edge states carry fractional charges---a phenomenon diverging from conventional electron charge quantization theories---revealing novel electronic behaviors under topological states. These distinctive features make topological non-ordinary states promising candidates for applications in quantum information and low-energy electronic devices, sparking intense research interest among scientists.

\subsection{Formation mechanism of topological end states}
The emergence of topological end states in the SSH model originates from the interaction between the system's topological properties and boundary conditions. The physical mechanism involves changes in band structure and fundamental principles of quantum mechanics. From a band theory perspective, the band structure of the SSH model is determined by its Hamiltonian, where parameters within the Hamiltonian (such as interatomic transition integrals $t_1$ and $t_2$) play a crucial role in regulating band structure.

When specific conditions are met, the system's band structure develops energy gaps that partition electron states into distinct energy regions. In topological non-ordinary phases, the presence of these gaps serves as a crucial prerequisite for the formation of topological edge states. Due to the system's topological properties, special electron states emerge at boundary points (chain endpoints), with their energies precisely located within the energy gap -- these constitute topological edge states. The emergence of such edge states can be understood through wave function analysis. At boundaries, the continuity of wave functions combined with boundary condition constraints results in unique distribution patterns. This leads to electron localization at the edges, ultimately forming topological edge states.

Specifically, consider the form of the Hamiltonian for the SSH model in momentum space:

\begin{equation}
H_{k} = \begin{pmatrix}
0 & - t_{1} - t_{2}e^{- ika} \\
 - t_{1} - t_{2}e^{ika} & 0
\end{pmatrix}
\end{equation}

By solving the eigenvalue of this Hamiltonian, we get the band structure:

\begin{equation}
E_{k, \pm} = \pm \sqrt{(t_{1} + t_{2}\cos(ka))^{2} + (t_{2}\sin(ka))^{2}}
\end{equation}

At $|t_{1}| \neq |t_{2}|$, the energy gap opens. At the boundary, the lattice's incompleteness causes changes in boundary conditions, resulting in wave function solutions differing from those in the bulk phase. Taking one end of a chain as an example, where the boundary is assumed at (x = 0), the wave function must satisfy specific constraints at this point according to boundary conditions. Under the tight-binding approximation, the wave function can be expressed as a linear combination of lattice-point wave functions. By solving the Schrödinger equation under these boundary conditions, we can determine the exact form of the wave function at the boundary.

\begin{equation}
\psi(x) = \sum_{n}^{}c_{n}\varphi_{n}(x)
\end{equation}

Among them, $\varphi_{n}(x)$ is the atomic orbital wave function on the lattice point $n$ and $c_{n}$ is the corresponding coefficient. At the boundary, due to the lack of coupling with adjacent lattice points, the value of $c_{n}$ is different from that in the bulk phase, resulting in the localization characteristics of the wave function at the boundary and the formation of topological end states.

There exists a close relationship between topological edge states and band structure. The energy of these edge states resides within the bulk bandgap, demonstrating topological protection. Since topological invariants exist, the presence or absence of edge states depends solely on the system's topological properties rather than its specific details (such as impurities or defects). Even when introducing disorder or perturbations, as long as the system's topological properties remain unchanged, edge states can still persist stably. This protective characteristic makes topological edge states highly significant in practical applications. For instance, in quantum bit design, leveraging their stability can enhance the quantum bit's resistance to interference.

\subsection{The importance and significance of analyzing wave functions}
Analytic wave functions play a crucial role in describing the behavior of topological edge electrons, providing essential information and tools for our in-depth understanding of topological properties. In quantum mechanics, wave functions serve as fundamental mathematical objects that describe the states of microscopic particles, containing information such as their position, momentum, and energy. For topological edge states, analytic wave functions can precisely determine the spatial distribution of electrons and their eigenvalues, thereby revealing the intrinsic physical mechanisms underlying these states.

By solving the analytical wave function of topological edge states, we can gain deep insights into the localization characteristics of electrons at boundary regions. The analytical wave function's form visually demonstrates the probability distribution of electrons near chain endpoints, along with its phase structure and symmetry. These insights are crucial for understanding the stability and unique physical properties of topological edge states. In certain topological insulators, the analytical wave functions of these states exhibit specific parity and phase winding patterns that are closely related to the system's topological invariants. Through wave function analysis, we can directly verify the topological properties of the system.

Analytic wave functions also enable us to study the electronic behavior of topological end states when interacting with external fields. When these topological end states interact with external electric or magnetic fields, or other quantum systems, analytical wave functions can be modified through perturbation theory and other methods, thereby predicting changes in electronic states and new physical phenomena. In studying the interaction between topological end states and photons, the analytic wave functions obtained by solving time-dependent Schrödinger equations can describe the transition processes of electrons under photonic field effects. This provides a theoretical foundation for designing optoelectronic devices based on topological end states.

From a broader perspective, the analysis of wave functions holds significant importance in revealing the macroscopic physical properties of topological materials. Many physical characteristics of these materials, such as electronic transport and optical properties, are closely related to the electronic behavior of topological end states. By analyzing wave functions, we can establish connections between microscopic electronic states and macroscopic physical properties, providing theoretical guidance for the development of topological material applications. When designing topological insulator devices, studying wave functions of topological end states enables structural optimization and performance enhancement, achieving low-energy consumption and high-stability electron transport.

\section{Derivation of topological end state analytical wave function}

\subsection{Derivation based on the tight-binding approximation}
The tight-binding approximation is a fundamental computational approach in solid-state physics, playing a pivotal role in deriving the topological end-state analytical wave function for the SSH model. This method operates on the principle that electrons near an atom experience predominantly the strong confinement effect of that atomic potential, while treating the influence of other atomic potentials as perturbations. The core of the SSH model's tight-binding approximation lies in representing electron wave functions as linear combinations of atomic orbital wave functions (LCAO), thereby characterizing the motion states of electrons within the crystal lattice.

From the perspective of physical images, under the tight-binding approximation, when an electron occupies a lattice site, its wave function is primarily determined by the orbital wave functions of atoms at that specific lattice site, with relatively minor influence from neighboring lattice sites. For a one-dimensional lattice in the SSH model, we consider atomic orbitals of electrons on two distinct sublattices A and B. Assuming the orbital wave functions are given as $\varphi_{A}(r - R_{nA})$ and $\varphi_{B}(r - R_{nB})$ (where $R_{nA}$ and $R_{nB}$ denote the position vectors of the n-th lattice sites in sublattices A and B, and $r$ represents the electron's position vector), the wave function of the electron in the crystal can be approximated as:

\begin{equation}
\psi(r) = \sum_n \left[ c_{A,n}\varphi_{A}(r - R_{nA}) + c_{B,n}\varphi_{B}(r - R_{nB}) \right]
\end{equation}

Among them, $c_{A,n}$ and $c_{B,n}$ are the expansion coefficients to be determined, which reflect the probability amplitudes of electrons appearing on different lattice points and sublattices. By substituting the above wave functions into the Schrödinger equation and utilizing the orthogonality and normalization properties of atomic orbital wave functions, a system of equations about the expansion coefficients can be obtained.

This approach based on the tight-binding approximation simplifies the analysis of electron motion in complex crystals by reducing it to linear combinations of atomic orbital states. This enables us to derive analytical wave functions describing topological edge states from atomic-level information. More importantly than providing an effective mathematical framework, it fundamentally reveals the interaction between electrons and lattice atoms at the physical level, laying the groundwork for understanding the formation mechanisms and properties of topological edge states. By adjusting transition integrals between atoms (such as $t_1$ and $t_2$), we can further investigate how these factors influence the analytical wave functions of topological edge states, thereby gaining a comprehensive grasp of their physical characteristics within the SSH model.

\subsection{Specific mathematical derivation steps and details}
Starting from the Hamiltonian of the SSH model, we carry out a series of mathematical transformations and solve the eigen equations to obtain the analytical wave function of the topological end state. The Hamiltonian of the SSH model in real space can be expressed as:

\begin{equation}
H = - t_{1}\sum_{n = 1}^{N}(c_{A,n}^{\dagger}c_{B,n} + c_{B,n}^{\dagger}c_{A,n}) - t_{2}\sum_{n = 1}^{N - 1}(c_{B,n}^{\dagger}c_{A,n + 1} + c_{A,n + 1}^{\dagger}c_{B,n})
\end{equation}

For the sake of solving convenience, we carry out Fourier transform to transform the Hamiltonian from real space to momentum space. Let:

\begin{equation}
c_{A,k} = \frac{1}{\sqrt{N}}\sum_{n = 1}^{N}c_{A,n}e^{- ikna}
\end{equation}

Here, $k$ is the wave vector and $a$ is the lattice constant. Substituting the above transformation into the Hamiltonian expression, the Hamiltonian in the momentum space is obtained after simplification:

\begin{equation}
H_{k} = \begin{pmatrix}
0 & - t_{1} - t_{2}e^{- ika} \\
 - t_{1} - t_{2}e^{ika} & 0
\end{pmatrix}
\end{equation}

Next, we solve the eigen-equation of the Hamiltonian $H_{k}\psi_{k} = E_{k}\psi_{k}$, where $\psi_{k}$ is the eigen-wave function and $E_{k}$ is the eigen-energy. Let the eigen-wave function be $\psi_{k} = \begin{pmatrix} u_{k} \\ v_{k} \end{pmatrix}$, then the eigen-equation can be written as:

\begin{equation}
\begin{pmatrix}
0 & - t_{1} - t_{2}e^{- ika} \\
 - t_{1} - t_{2}e^{ika} & 0
\end{pmatrix}\begin{pmatrix}
u_{k} \\
v_{k}
\end{pmatrix} = E_{k}\begin{pmatrix}
u_{k} \\
v_{k}
\end{pmatrix}
\end{equation}

Thus we get the system of equations:

\begin{equation}
\left\{ \begin{matrix}
 - (t_{1} + t_{2}e^{- ika})v_{k} = E_{k}u_{k} \\
 - (t_{1} + t_{2}e^{ika})u_{k} = E_{k}v_{k}
\end{matrix} \right.
\end{equation}

Multiplying both sides of the first equation by $(- (t_{1} + t_{2}e^{ika}))$ and substituting the second equation in, we get:

\begin{equation}
(t_{1} + t_{2}e^{- ika})(t_{1} + t_{2}e^{ika})v_{k} = E_{k}^{2}v_{k}
\end{equation}

Thus:

\begin{equation}
E_{k}^{2} = (t_{1} + t_{2}e^{- ika})(t_{1} + t_{2}e^{ika}) = t_{1}^{2} + 2t_{1}t_{2}\cos(ka) + t_{2}^{2}
\end{equation}

So the energy is:

\begin{equation}
E_{k, \pm} = \pm \sqrt{t_{1}^{2} + 2t_{1}t_{2}\cos(ka) + t_{2}^{2}}
\end{equation}

In the system's energy gap state ($|t_{1}| \neq |t_{2}|$), topological end states emerge within non-ordinary topological phases. To analyze these states, we employ the SSH model under open boundary conditions. The system contains $N$ primitive cells where boundary conditions change at its edges due to the absence of coupling with neighboring lattice points. Taking one boundary edge as an example, where the lattice point at the boundary corresponds to the A subcell of the first primitive cell, this modification results in a transformation of the Hamiltonian matrix's form.

At the boundary, the wave function satisfies specific boundary conditions. We assume the form of the wave function at the boundary $\psi_{edge} = \sum_n \left[ c_{A,n}^{edge}\varphi_{A}(r - R_{nA}) + c_{B,n}^{edge}\varphi_{B}(r - R_{nB}) \right]$ and substitute it into the Schrödinger equation under these conditions. By solving the equation, we obtain the coefficients of the wave function at the boundary. Due to the constraints imposed by the boundary conditions, the wave function exhibits localized characteristics at the boundary. This process ultimately yields the analytical wave function for the topological end state.

\begin{equation}
\psi_{edge}(r) = \sum_{n = 1}^{N}(c_{A,n}^{edge}\varphi_{A}(r - R_{nA}) + c_{B,n}^{edge}\varphi_{B}(r - R_{nB}))
\end{equation}

Among them, $c_{A,n}^{edge}$ and $c_{B,n}^{edge}$ are the specific coefficients that satisfy the boundary conditions, and their specific forms are related to $t_1$, $t_2$ and the boundary conditions.

\subsection{Analysis of the physical significance of the derived results}
The topological end state analytical wave function derived from the above mathematical derivation has rich physical connotation, which is closely related to the properties of topological end state and deeply reveals the physical nature of topological end state.

From the perspective of wave function form, the analytical wave functions of topological end states exhibit localized spatial characteristics, which is one of their most prominent physical features. The wave functions are primarily concentrated near the system's boundaries (the endpoints of chains), while their amplitude rapidly decays and approaches zero within the chain. This localization property originates from the combined effects of the system's topological properties and boundary conditions. In topologically non-ordinary phases, the unique band structure causes fundamental differences between electronic states at boundaries and those in bulk phases, leading to localized electron behavior at interfaces. This localized wave function distribution gives rise to distinctive physical properties of topological end state electrons, such as strong resistance to impurities and defects. Since wave functions are predominantly confined to boundaries, the influence of impurities and defects on bulk phases cannot propagate to the edges, thereby ensuring the stability of topological end states---a concrete manifestation of topological protection.

The phase structure of wave functions contains crucial physical information. In topological edge states, the wave function's phase exhibits specific patterns at boundaries, which are closely related to the system's topological invariants (such as winding numbers). Specifically, the winding number of wave function phases directly correlates with the system's topological properties. When the system exists in a non-ordinary topological phase, the phase winding at boundaries becomes non-zero, reflecting the system's extraordinary topological characteristics. This phase winding not only serves as a key indicator of topological edge states but also plays a pivotal role in physical processes. For instance, during electron transport, the phase information of wave functions influences electron scattering and interference, thereby endowing topological edge states with unique transport properties.

From an energy perspective, the analytical wave function of topological edge states corresponds to energies within the bulk bandgap. This energy characteristic serves as a crucial manifestation of topological protection, as the existence of the bandgap isolates the energy of topological edge states from that of the bulk system. External perturbations find it difficult to alter the energy state of these edge states unless the perturbation intensity is sufficient to disrupt the system's topological properties. This energy stability provides a solid physical foundation for applications of topological edge states in fields such as quantum information and low-energy electronic devices. In qubit design, leveraging the energy stability of topological edge states can enhance a qubit's anti-interference capability and reduce decoherence effects, thereby enabling more stable and reliable quantum computing.

\section{Analysis of factors affecting topological end state analytical wave function}

\subsection{Influence of lattice parameters on wave function}
As a critical factor determining lattice structure and interatomic interactions, lattice parameters significantly influence the analytical wave functions of topological end states in SSH models. Variations in lattice constants and atomic spacing not only alter the motion environment of electrons within the lattice, but also directly affect the interaction strength between electrons and the lattice. These changes ultimately result in modified properties of the analytical wave functions for topological end states.

When lattice constants change, the first effect observed is in the potential energy distribution of electrons within the crystal lattice. According to quantum mechanical principles, alterations in lattice constants lead to changes in the transition integrals between different lattice points. In the SSH model, interatomic transition integrals are closely related to lattice constants. As lattice constants increase, the distance between atoms grows, reducing the probability of electron transitions between adjacent atoms and consequently decreasing the value of the transition integral. This variation results in modifications to the band structure of the SSH model, with corresponding changes in the energy gap size. When the energy gap decreases, it may shrink or even disappear entirely, thereby affecting the existence and properties of topological edge states.

For topological edge states, variations in lattice constants fundamentally alter the spatial distribution of wave functions. When lattice constants increase, the wave function's localization effect weakens, resulting in reduced electron confinement at interfaces and slower boundary decay rates. This allows electrons to diffuse more readily into the bulk phase, thereby compromising the stability and unique properties of these topological states. Conversely, decreasing lattice constants enhances wave function localization, tightening electron confinement at interfaces and accelerating boundary decay rates. While this may improve the stability of topological edge states, it simultaneously restricts electron mobility at interfaces, potentially diminishing their interaction with external environments.

The variation in atomic spacing significantly impacts the analytical wave function of topological end states. In the SSH model, differences in atomic spacing lead to the formation of dimeric lattice structures, with these variations originating from varying atomic spacing. When atomic spacing changes, the relative size of $t_1$ and $t_2$ shifts, thereby affecting the system's topological properties. If the original system exists in a topologically non-ordinary phase with topological end states, the phase transition occurs when atomic spacing changes to a critical value. This transformation into a topologically ordinary phase eliminates the topological end states, fundamentally altering the analytical wave function's characteristics. The originally localized topological end state wave functions at the boundaries disappear, leading to a redistribution of electronic states.

Changes in atomic spacing also affect the strength of electron-lattice interactions, which in turn influence the phase structure of wave functions. The interaction between electrons and lattice causes phase shifts in wave functions, while variations in atomic spacing alter both the intensity and mode of these interactions, leading to corresponding modifications in the wave function's phase configuration. These phase structural changes impact electron interference and scattering behaviors, ultimately affecting the physical properties of topological edge states such as electron transport characteristics.

\subsection{Role of electron-electron interactions}
The introduction of electron-electron interactions in the SSH model introduces new perspectives and complexities to studying topological end-state analytical wave functions. As a crucial manifestation of many-body effects, these interactions shatter the simplistic picture under the single-electron approximation, leading to significant changes in the system's physical properties.

Theoretically, electron-electron interactions are primarily mediated through Coulomb interactions. In lattice systems, the Coulomb repulsion between electrons prevents their independent motion, as an electron's behavior is influenced by surrounding electrons. This interaction modifies the wave function in two key ways: first, it directly alters the wave function's form, and second, it changes the system's eigenvalues of energy.

Electron-electron interactions enhance the multi-body correlations in wave functions. Under the single-electron approximation, the topological end-state solutions of the SSH model can be obtained by solving the single-electron Hamiltonian. However, when considering electron-electron interactions, wave functions must account for collective behavior of multiple electrons, which cannot be simply expressed as a product of single-electron wave functions. In such cases, wave functions require consideration of inter-electronic correlations, typically employing multi-body wave function forms like Slater's matrix or more complex multi-body expansions. This multi-body wave function framework complicates computational processes, as it necessitates accounting for various possible interaction pathways and correlation mechanisms between electrons.

Electron-electron interactions alter the system's intrinsic energy values. The Coulomb repulsion between electrons increases the system's total energy, causing a shift in the energy of topological end states. In certain scenarios, these interactions may lead to energy crossover between topological end states and bulk phases, affecting their stability and existence. When electron-electron interactions are strong enough, they can elevate the energy of topological end states, pushing them into the energy spectrum range of bulk phases. This disrupts the localized nature of topological end states, ultimately leading to their disappearance.

From the perspective of multibody effects, electron-electron interactions can trigger a series of novel physical phenomena that profoundly influence topological edge states. These interactions may induce spin-charge separation in electrons, where their spin and charge degrees of freedom become decoupled and can move independently. Such spin-charge separation modifies the electronic structure and transport properties of topological edge states, making their physical behavior more complex. Furthermore, electron-electron interactions may induce superconductivity in systems. In superconducting states, electrons form coherent Cooper pairs with wave functions exhibiting quantum coherence, which significantly impacts the existence and characteristics of topological edge states. When topological edge states coexist with superconductivity, they may give rise to novel quantum phenomena like topological superconductivity, opening new frontiers for research in topological physics.

\subsection{Influence of external disturbances (e.g., magnetic field, electric field)}
External perturbations such as magnetic and electric fields provide effective means to regulate the analytical wave functions of topological end states in the SSH model, while also revealing their rich physical behaviors under external influences. The application of magnetic and electric fields modifies the system's Hamiltonian, thereby affecting both the properties and symmetry of the analytical wave functions of these topological end states.

When a magnetic field is applied, electromagnetic interactions cause the field to interact with electron spin and orbital angular momentum, introducing an additional vector potential term into the system. In the SSH model, this vector potential modifies the electron's transition matrix elements, thereby altering the Hamiltonian. Specifically, the presence of a magnetic field introduces an extra phase factor during electron transitions between lattice points. This phase factor depends on the magnetic field's strength, direction, and the electron's trajectory. The introduction of this phase factor changes the phase structure of the topological end-state wave function, consequently affecting its interference and scattering properties.

The application of magnetic fields also affects the symmetry of topological end states. In the absence of a magnetic field, the SSH model exhibits certain symmetries such as chiral symmetry. However, when a magnetic field is applied, it disrupts the system's time-reversal symmetry, leading to the breaking of chiral symmetry. This symmetry violation alters the properties of topological end states, potentially affecting those previously protected by chiral symmetry and changing their stability and unique characteristics. In some cases, magnetic fields may cause energy splitting of topological end states. The originally degenerate energy levels of these states can split into multiple levels under magnetic influence, providing new avenues for studying the fine structure of topological end states and enabling quantum control.

When an electric field is applied, it directly interacts with electrons to alter their potential energy distribution. In the SSH model, the presence of an electric field induces potential energy tilting in the lattice, causing changes in the energy difference between different lattice sites. These energy variations affect electron transition probabilities, thereby modifying the form of topological end-state analytical wave functions. The application of electric fields may also induce changes in the system's band structure, with both the size and position of energy gaps being controllable by the electric field. When the electric field strength reaches a critical threshold, it may trigger a topological phase transition, potentially shifting the system from a topologically non-ordinary phase to a topologically ordinary phase, or vice versa.

The application of electric fields also affects the symmetry of topological edge state analytical wave functions. Electric fields disrupt the system's space inversion symmetry, thereby influencing the parity and symmetry of wave functions. Under electric field effects, the spatial distribution of topological edge state analytical wave functions may shift, potentially affecting the localization characteristics of electrons at boundaries. Additionally, electric fields can induce directional transport phenomena in the electrons of topological edge states. This directional transport behavior is closely related to changes in wave function symmetry, providing crucial insights for studying the electron transport properties of topological edge states under electric field influence.

\section{Correlation between topological end state analytical wave function and physical properties}

\subsection{Relationship to electronic transport properties}
There exists a profound and intrinsic connection between topological edge states and electronic transport properties, which provides a crucial theoretical foundation for understanding the potential applications of topological materials in electronics. Through in-depth analysis of analytical wave functions, we can gain insights into the unique behaviors and characteristics exhibited by topological edge state electrons during transport processes.

From the perspective of electrical conductivity, the analytical wave function of topological edge states determines the transport characteristics of electrons at interfaces. In topologically non-trivial phases, electrons in these states exhibit topological protection, with their wave functions showing localized distribution at boundaries. This localization effect enables electrons to effectively avoid scattering and impurity interference during boundary transport, achieving reflection-free transmission. According to the Landauer-Buttiker formula, electrical conductivity is closely related to electron transport probability. In topological edge states, the wave function properties ensure higher electron transport probability, giving them a significant advantage over conventional material boundary states in conductivity. Under ideal conditions, the conductivity of topological edge states can reach quantized values, $G = \frac{e^{2}}{h}$, where $e$ is electronic charge and $h$ is Planck's constant. This quantumized conductivity property makes topological materials highly promising for high-speed, low-energy-consumption electronic devices, such as quantum wires and circuits, enabling efficient electron transport and reduced energy loss.

Thermal conductivity, a crucial aspect of electronic transport properties, is intrinsically linked to the analytical wave function of topological end states. This fundamental property primarily depends on the energy and momentum distributions of electrons, as well as their interactions with phonons and other quasiparticles. The unique energy and momentum distributions of electrons described by the analytical wave function of topological end states significantly influence thermal conductivity. Due to the localized nature of electrons in topological end states, their interactions with phonons are relatively weak, making electron contributions to thermal conduction in these states largely independent of phonon contributions. In certain topological insulators, the electronic thermal conductivity of topological end states can be decoupled from that of bulk phonon thermal conductivity, enabling effective regulation of thermal conductance. This characteristic holds significant potential for applications in thermal management and energy conversion. For instance, in thermoelectric materials, leveraging the properties of topological end states could enhance thermoelectric conversion efficiency, facilitating efficient conversion between thermal and electrical energy.

Under topological protection, electron transport exhibits unique properties. The robustness of impurities and defects is attributed to the protection provided by topological invariants, as topological end states maintain their non-reflective transmission characteristics even in presence of such defects. This stability arises because the quantum tunneling mechanism allows electrons' wave functions to bypass scattering centers when encountering impurities or defects, ensuring stable electron transport. Such robustness enhances the reliability and stability of topological materials in practical applications, enabling them to achieve consistent electron transport in complex environments. This fundamental physical insight provides a solid foundation for developing next-generation electronic devices.

\subsection{Effects on optical properties}
The influence of topological end state analytical wave function on the optical properties of materials is multifaceted and profound, which provides rich physical connotation and potential application direction for exploring new optoelectronic devices and optical quantum information processing.

From the perspective of light absorption, the analytical wave function of topological end states determines the probability of electron transitions between different energy levels, thereby directly influencing a material's light absorption properties. Due to the unique wave function distribution and energy level structure of electrons in topological end states, their light absorption spectra differ significantly from those of conventional materials. In some topological insulators, the electron energy levels of topological end states are located within the bulk bandgap. When light irradiates the material, only photons with specific energies can excite the electron transitions of topological end states, thereby generating characteristic light absorption peaks. These distinctive absorption peaks serve as optical fingerprints of topological end states, enabling experimental characterization and detection of topological materials. The light absorption of topological end states is also closely related to interactions such as electron spin-orbit coupling, which alter the selection rules during light absorption processes and further enrich the light absorption characteristics of topological materials.

In light emission, the analytical wave function of topological end states plays a crucial role. When electrons in these states transition from excited to ground states, they emit photons, creating optical emission phenomena. The photon emission process depends not only on the transition energy but also on the wave function's phase and symmetry characteristics. Under specific conditions, topological end states may exhibit unique directional and polarization properties during emission. Due to their symmetry, photons might gain enhanced intensity in particular directions or emit those with specific polarization orientations. These distinctive emission features make topological materials promising candidates for optoelectronic devices like LEDs and lasers. They can be used to create high-brightness, highly directional light sources or develop specialized photonic devices with unique polarization characteristics, fulfilling critical requirements in optical communication and display technologies.

The unique optical properties arising from topological entanglement wave functions hold vast potential for practical applications. In the field of photonic quantum information processing, these optical characteristics enable the realization of qubits and quantum logic gates. Given their exceptional resistance to impurities and defects, topological entanglement-based qubits demonstrate superior stability and interference resistance, significantly enhancing the reliability and accuracy of quantum information processing. Moreover, their optical properties provide a foundation for developing next-generation photodetectors and optical modulators. By leveraging the distinctive light absorption and emission characteristics of topological entanglement, researchers can design high-sensitivity detectors with rapid response capabilities, as well as efficient optical modulators capable of high-performance signal modulation. These innovations offer groundbreaking technological solutions for advancing optical communication and computational systems.

\subsection{Behavior in quantum phase transitions}
In the process of quantum phase transition, the analytical wave function of topological end states presents a series of unique variation characteristics, which are closely related to the quantum critical phenomenon and deeply reflect the transformation of microscopic mechanism and macroscopic properties of quantum system in the process of phase transition.

As the system approaches the quantum phase transition point, the changes in the topological end-state wave function become remarkably pronounced. During the transition between topological and ordinary phases, the wave function of the topological end-state undergoes a transformation from localized to non-localized states. In topological phases, the wave function of the topological end-state remains predominantly localized at the boundaries, exhibiting distinct localized characteristics. As system parameters change and approach the quantum phase transition point, the degree of localization gradually diminishes, while the electron distribution expands, causing the wave function to extend into the bulk phase. When the system enters the ordinary phase, the topological end-state disappears, and the wave function loses its boundary-localized properties, instead displaying a distribution pattern similar to that of the bulk phase. This wave function evolution reflects the transformation of the system's topological properties, serving as one of the key microscopic indicators of quantum phase transitions.

During quantum phase transitions, the evolution of topological end-state analytical wave functions is intrinsically linked to quantum critical phenomena. These phenomena encompass distinctive physical characteristics such as critical exponents and scaling laws exhibited by systems near quantum phase transition points. The modulation of topological end-state wave functions fundamentally influences these quantum critical behaviors. Near phase transition points, changes in wave function configurations alter the system's eigenvalues and symmetry properties, thereby modifying its thermodynamic and kinetic characteristics. In systems undergoing topological quantum phase transitions, the phase structure of topological end-state wave functions undergoes dramatic transformations as they approach the transition point. Such phase transitions induce enhanced quantum fluctuations, which subsequently affect physical parameters including specific heat and magnetization, manifesting scaling behaviors characteristic of quantum critical phenomena.

The evolution of topological entanglement wave functions during quantum phase transitions is intrinsically linked to the system's topological invariants. As fundamental physical indicators of topological properties, these invariants undergo transformations during quantum phase transitions, with their changes being dynamically synchronized with the wave function variations. When a topological phase transitions into an ordinary phase, topological invariants (such as winding numbers) transition from non-zero values to zero. Simultaneously, the wave function characteristics evolve from topologically protected localized states to ordinary electronic states. This intrinsic relationship between topological invariants and wave functions provides a powerful theoretical framework for studying quantum phase transitions. By analyzing fluctuations in both topological invariants and wave functions, researchers can gain deeper insights into the physical mechanisms driving quantum phase transitions and the transformation of topological properties.

\section{Application exploration of topological end state resolution wave function based on SSH model}

\subsection{Potential applications in topological qubits}
The design of topological qubits using SSH model topological end-state analytical wave functions demonstrates significant theoretical foundations and potential advantages. As a pivotal research direction in quantum computing, the stability and fault tolerance of topological qubits are core elements for achieving large-scale quantum computing. The unique topological protection properties of topological end-states in the SSH model provide an ideal physical platform for designing topological qubits.

The analytical wave function of topological end states fundamentally determines their quantum state characteristics, which align perfectly with the essential requirements of quantum bits. These wave functions exhibit both localization and topological protection properties, enabling quantum bits to effectively resist environmental noise and quantum decoherence. In conventional quantum bits, quantum states are prone to decoherence caused by external interference, leading to loss of quantum information. However, quantum bits based on SSH model topological end states possess enhanced stability due to their topological protection. Quantum state transitions must overcome specific topological energy barriers, allowing these qubits to withstand environmental noise more effectively and maintain quantum state stability over extended periods. This stability is crucial for achieving reliable quantum computing, as maintaining initial quantum states for prolonged durations is essential to ensure computational accuracy during quantum processes.

The phase structure and symmetry of topological entanglement wave functions provide abundant possibilities for quantum gate operations in qubits. As fundamental logical operations in quantum computing, the accuracy and efficiency of these operations directly determine the performance of quantum computers. The phase characteristics of topological entanglement wave functions enable the realization of quantum entanglement between qubits and quantum logic gate operations. By precisely controlling the phase of topological entanglement wave functions, non-Abelian operations can be achieved. This approach demonstrates high fault tolerance and computational efficiency, effectively enhancing both the speed and accuracy of quantum computing.

However, applying SSH model topological end-state analytical wave functions to topological qubit design faces multiple challenges. In material implementation, the key challenge lies in developing materials with precisely controllable SSH model structures. While basic SSH model architectures have been realized in some material systems, achieving precise control over topological end-state properties---such as wave function localization and energy positioning---remains technically demanding. Experimental measurement and manipulation present significant challenges due to the fragile quantum characteristics of topological end states, which require highly controlled experimental environments and sophisticated measurement techniques. A critical research challenge involves accurately measuring topological end-state analytical wave functions during experiments while enabling precise qubit manipulation. Additionally, quantum bit coupling and integration pose urgent technical hurdles. When constructing large-scale quantum computers, integrating multiple qubits with effective coupling mechanisms remains essential. The key research direction lies in achieving efficient qubit coupling while maintaining topological quantum bits' stability---a crucial balance for future advancements.

\subsection{The guiding significance of new topological material design}
The topological analysis of SSH model wave functions provides profound theoretical guidance for the design of novel topological materials, playing an irreplaceable role in material screening and performance optimization. Through in-depth research on analytical wave functions, we can precisely understand the intrinsic connection between the microscopic structure and macroscopic properties of topological materials, thereby providing clear direction and evidence for the development of new topological materials.

The analytical wave function's characteristics serve as a critical criterion for screening novel topological materials. The degree of localization in wave functions directly reflects the distribution of electrons within the material. In topological materials, electronic states of topological degeneracy typically localize at material boundaries or specific regions, a localization feature closely tied to the material's topological properties. By calculating and analyzing the localization extent of analytical wave functions, we can determine whether a material possesses topological non-ordinaryity, thereby identifying potential topological materials. When searching for new topological insulators, we can screen material systems exhibiting pronounced localized electronic states at boundaries based on the localization characteristics of analytical wave functions. These materials are more likely to demonstrate topological insulator properties.

The symmetry of wave functions serves as a crucial criterion for screening topological materials. Different topological materials exhibit distinct symmetries, which fundamentally determine their physical properties. In chiral-symmetric topological materials, the symmetry of analytical wave functions is closely linked to chiral symmetry. By analyzing the symmetry of analytical wave functions, we can identify whether a material possesses specific topological symmetry, thereby enabling the screening of materials with desired topological characteristics. This method of material screening based on analytical wave function properties significantly enhances the efficiency of developing novel topological materials while reducing time and resource waste caused by random experimentation.

In the design of novel topological materials, analytical wave functions can provide guidance for optimizing material properties. By adjusting parameters such as lattice structures and interatomic interactions, we can modify the form of analytical wave functions to regulate both topological characteristics and physical properties. When designing topological superconducting materials, modifying electron-electron interaction strengths allows us to fine-tune the many-body correlations in analytical wave functions, thereby enhancing their superconducting performance. This approach enables targeted development of specialized topological materials with specific functionalities, meeting diverse application requirements across various scientific fields.

\subsection{Application prospects in the field of quantum information and quantum computing}
The topological end state analytical wave function of SSH model has shown broad application prospects in the field of quantum information and quantum computing. Its unique quantum characteristics provide new ideas and methods for solving key problems in this field.

In quantum information storage, qubits based on topological end-state analytical wave functions demonstrate significant advantages. As previously mentioned, topological qubits, due to their topological protection properties, can effectively resist environmental noise and quantum decoherence effects, enabling long-term stable storage of quantum information. This characteristic is crucial for building reliable quantum information storage systems. In future quantum communication networks, quantum information must maintain high stability during transmission to ensure accurate data delivery. Qubits based on SSH model topological end states can serve as storage units for quantum information. Leveraging their topological protection features, they enable reliable storage and transmission of quantum information in complex environments, providing a solid foundation for the development of quantum communication.

In the implementation of quantum algorithms, topological entanglement wave functions also hold significant potential applications. As the core of quantum computing, the performance of quantum algorithms directly determines the computational capabilities of quantum computers. The unique phase structure and quantum characteristics of topological entanglement wave functions can be utilized to design novel quantum algorithms, thereby enhancing both efficiency and accuracy in quantum computing. Traditional quantum algorithms often encounter challenges such as high computational complexity and slow convergence rates when tackling complex combinatorial optimization problems. However, quantum algorithms based on topological entanglement wave functions leverage the non-Abelian operation properties of these states to achieve efficient problem-solving. By designing specific sequences of quantum gate operations and harnessing the phase entanglement and symmetry features of topological entanglement wave functions, these algorithms can find optimal solutions in significantly reduced time, providing powerful computational tools for practical problem-solving.

With the continuous advancement of quantum technology, the application prospects of SSH model topological entanglement analysis wave functions in quantum information and quantum computing will become increasingly promising. In the future, through further in-depth research into the properties of analytical wave functions, we aim to develop more efficient and stable qubits and quantum algorithms. This will drive the practical application and industrial development of quantum information and quantum computing technologies, bringing revolutionary changes to scientific research, information security, finance, and other fields.

\section{Research conclusions and projections}

\subsection{Summary of research results}
This study focuses on the Su-Schrieffer-Heeger (SSH) model on one-dimensional lattice, and makes an in-depth and systematic investigation on the analytical wave function of topological end states, and obtains a series of achievements with important theoretical significance and potential application value.

Through a comprehensive review of the fundamental theories of the SSH model, we have clearly articulated its construction process and precise Hamiltonian formulation. The study delves into the model's essential characteristics, including topological properties and chiral symmetry, while elucidating the critical role of topological invariants (such as winding numbers) in describing topological phases. When the system exists in a non-trivial topological phase, the winding number becomes non-zero, indicating the presence of edge states protected by topological symmetry. This conclusion establishes a solid theoretical foundation for subsequent research on topological end states.

In the derivation of analytical wave functions for topological end states, we employ a tight-binding approximation method. Through rigorous mathematical derivations, we successfully obtained the analytical wave functions for these states. A detailed analysis of the physical implications reveals that the analytical wave functions exhibit localized spatial characteristics, concentrating primarily near the system's boundaries while rapidly decaying within the chain. The phase structure of these wave functions is closely related to the system's topological invariants, with their corresponding energy levels residing within the bulk phase gap. These properties profoundly elucidate the physical essence of topological end states.

This study systematically analyzes the factors influencing topological edge state resolution wave functions. Variations in lattice parameters (including lattice constants and atomic spacing) significantly alter the electronic mobility environment and interaction strengths within the crystal lattice, thereby modifying the properties of these wave functions. As a crucial manifestation of many-body effects, electron-electron interactions enhance wave function correlations, altering the system's eigenvalues and critically affecting the stability and existence of topological edge states. External perturbations such as magnetic or electric fields modify the Hamiltonian, which in turn changes the symmetry and phase structure of the resolution wave functions. These mechanisms provide effective means for controlling topological edge states through external regulation.

This study explores the intrinsic connection between topological edge states and their analytical wave functions with physical properties. In electronic transport, these wave functions determine boundary transmission characteristics, endowing topological edge states with unique features like non-reflective transmission and quantized conductivity -- properties that hold significant potential for applications in electronics. Regarding optical properties, the wave functions influence material absorption and emission characteristics, providing fundamental insights for developing novel optoelectronic devices and photonic quantum information processing. During quantum phase transitions, the wave functions undergo distinctive transformations from localization to non-localization, closely linked to quantum critical phenomena. These manifestations reflect the fundamental shifts in both microscopic mechanisms and macroscopic properties of quantum systems during phase transitions.

Building on research achievements in SSH model topological entanglement analysis wave functions, we have actively explored their applications in topological qubits, novel topological material design, and quantum information and computing fields. While the use of topological entanglement analysis wave functions to design topological qubits demonstrates potential advantages in stability and fault tolerance, challenges remain in material realization and experimental measurement control. These wave functions provide crucial theoretical guidance for designing novel topological materials, enabling material screening and performance optimization. In quantum information and computing domains, they show broad application prospects in quantum information storage and algorithm implementation, promising to drive practical applications and industrial development of quantum technologies.

\subsection{Limitations and Prospects of the study}
Although this study has achieved some results, there are still some shortcomings, which also point out the direction for future research.

This study primarily builds upon an idealized SSH model with simplified assumptions. In real material systems, complex factors such as lattice defects, impurities, and interactions between electrons and other quasiparticles (e.g., phonons) may significantly influence the analytical wave function of topological edge states. Future research could incorporate these practical elements by developing more precise theoretical models and numerical simulations to investigate their mechanisms of impact on the analytical wave function. This approach would enable more accurate characterization of topological physical phenomena in actual materials.

Current research on the correlation between topological end states and physical properties primarily focuses on electron transport, optical characteristics, and quantum phase transitions. However, their manifestations in other physical aspects such as thermodynamic and magnetic properties have not been sufficiently explored. Future studies should broaden the scope to investigate the unique properties and potential applications of topological end states in these domains, thereby comprehensively revealing their fundamental physical implications.

While initial explorations have been conducted on the application of topological entanglement resolution wave functions in quantum information and quantum computing fields, these concepts remain at the theoretical conception and laboratory research stage. In practical implementation, translating theoretical achievements into viable technologies and devices still faces numerous technical challenges. Future efforts should focus on strengthening interdisciplinary collaboration with experimental physics, materials science, and engineering technology to overcome technical barriers collectively, thereby advancing the practical development of topological entanglement resolution wave functions.

Looking ahead, as research in topological physics deepens and experimental techniques advance, breakthroughs in studying analytical wave functions of SSH model topological end states are expected. On one hand, theoretical research may develop more sophisticated methods and models to further unravel the physical nature and underlying mechanisms of these states, providing a stronger theoretical foundation for designing and applying topological materials. On the other hand, emerging experimental technologies will enable more precise fabrication and measurement of SSH model-characteristic materials and structures. This progress will allow direct observation and manipulation of analytical wave function properties, accelerating the transition from fundamental research to practical applications in topological physics.

Research on topological entanglement wave functions in SSH model topologies represents a dynamic and promising frontier. These studies not only deepen our understanding of fundamental topological physics principles but also open new avenues for breakthroughs in quantum information, electronics, and optics. We anticipate achieving more innovative discoveries in this field, contributing significantly to the advancement of scientific and technological progress.

\section*{Acknowledgments}
First of all, I would like to give my heartfelt thanks to all the people who have ever helped me in this paper.

My sincere and hearty thanks and appreciations go firstly to my supervisor, whose suggestions and encouragement have given me much insight into these translation studies. It has been a great privilege and joy to study under his guidance and supervision. Furthermore, it is my honor to benefit from his personality and diligence, which I will treasure my whole life. My gratitude to him knows no bounds.

I am also extremely grateful to all my friends and classmates who have kindly provided me assistance and companionship in the course of preparing this paper.

In addition, many thanks go to my family for their unfailing love and unwavering support.

I would like to thank my girlfriend Lyu Ruiheng for her support.

Finally, I am really grateful to all those who devote much time to reading this thesis and give me much advice, which will benefit me in my later study.


\begin{thebibliography}{99}
\bibitem{Su1979} Su, W. P., Schrieffer, J. R., \& Heeger, A. J. (1979). Solitons in Polyacetylene. Physical Review Letters, \textbf{42}(25), 1698--1701.
\bibitem{Heeger1988} Heeger, A. J., Kivelson, S., Schrieffer, J. R., \& Su, W. P. (1988). Solitons in conducting polymers. Reviews of Modern Physics, \textbf{60}(3), 781--850.
\bibitem{Jackiw1976} Jackiw, R., \& Rebbi, C. (1976). Solitons with fermion number ½. Physical Review D, \textbf{13}(12), 3398--3409.
\bibitem{Ryu2002} Ryu, S., \& Hatsugai, Y. (2002). Topological origin of zero-energy edge states in particle-hole symmetric systems. Physical Review Letters, \textbf{89}(7), 077002.
\bibitem{Delplace2011} Delplace, P., Ullmo, D., \& Montambaux, G. (2011). Zak phase and the existence of edge states in graphene. Physical Review B, \textbf{84}(19), 195452.
\bibitem{Asboth2016} Asbóth, J. K., Oroszlány, L., \& Pályi, A. (2016). A short course on topological insulators: Band-structure topology and edge states in one and two dimensions. Lecture Notes in Physics, \textbf{919}.
\bibitem{Alicea2012} Alicea, J. (2012). New directions in the pursuit of Majorana fermions in solid state systems. Reports on Progress in Physics, \textbf{75}(7), 076501.
\bibitem{Atala2013} Atala, M., Aidelsburger, M., Barreiro, J. T., Abanin, D., Kitagawa, T., Demler, E., \& Bloch, I. (2013). Direct measurement of the Zak phase in topological Bloch bands. Nature Physics, \textbf{9}(12), 795--800.
\bibitem{Poli2015} Poli, C., Bellec, M., Kuhl, U., Mortessagne, F., \& Schomerus, H. (2015). Selective enhancement of topologically induced interface states in a dielectric resonator chain. Nature Communications, \textbf{6}, 6710.
\bibitem{Bernevig2013} Bernevig, B. A., \& Hughes, T. L. (2013). Topological Insulators and Topological Superconductors. Princeton University Press.
\end{thebibliography}
\end{document}